\documentclass[conference]{IEEEtran}
\IEEEoverridecommandlockouts
\usepackage[pdftex]{graphicx}
\usepackage{lscape}
\usepackage{nidanfloat}
\usepackage{subfigure}

\title{Usefulness of Instructor Annotations on Flipped Learning Preparation Video System}
\author{
  \IEEEauthorblockN{
    Shintaro Uchiyama\IEEEauthorrefmark{1}, Hayato Okumoto\IEEEauthorrefmark{1}$^1$\thanks{$^1$Present affiliation is TwoGate Inc., Shinagawa, Tokyo, Japan; Email: okumoto@twogate.com}, Mitsuo Yoshida\IEEEauthorrefmark{1}, Yuko Ichikawa\IEEEauthorrefmark{3} and Kyoji Umemura\IEEEauthorrefmark{1}}
  \IEEEauthorblockA{\IEEEauthorrefmark{1}Department of Computer Science and Engineering\\
  Toyohashi University of Technology \\
  Toyohashi, Aichi, Japan\\
  Email: s183313@edu.tut.ac.jp, h153317@edu.tut.ac.jp, yoshida@cs.tut.ac.jp, umemura@tut.jp
  }
  \IEEEauthorblockA{\IEEEauthorrefmark{3}General Education Department\\
  National Institute of Technology,
  Tokyo College\\
  Hachioji, Tokyo, Japan\\
  Email: yuko@tokyo-ct.ac.jp}
}
\begin{document}
\maketitle

\IEEEpubidadjcol
\begin{abstract}
  Flipped learning is a method that flips in/out class activities to make lectures \mbox{learner-centered}.
  In flipped learning, comments from learners on preparation material are  useful information for instructors to consider before deciding in-class topics.
  Thus, we arrive at the notion that receiving comments from instructors will be effective for learners watching the video.
  By including annotations from instructors, we propose to improve the quality of content for learners and thus enhance learners' motivation and study satisfaction.
  To achieve this, we introduced ``Steering Mark,'' a tool that enables learners to easily grasp the overall structure of a video, to the video learning system.
  We examined the effectiveness and influence of Steering Mark through an experiment with 34 undergraduate learners.
  As a result, Steering Mark was found to be useful in improving the quality of video content for learners.
\end{abstract}

\section{Introduction}
Recently, flipped learning has become a popular learning method~\cite{SamsAaronJonathan2012}~\cite{Bishop}~\cite{Thai2017}.
Traditionally, instructors teach learners in the classroom, and learners review and apply this knowledge to homework outside the classroom.
Flipped learning flips these in/out classroom activities.
Learners learn by watching videos or using other material outside the classroom, and then they review and apply this knowledge in the classroom.
This method improves learning interest because it allows learners to take in information at their own pace during the out-of-classroom preparation time, as well as enables the learner and instructor to spend more time on communicating and discussing topics in the classroom. Therefore, flipped learning turns traditional lectures into learner-centered education.

In several applications of flipped learning~\cite{Abeysekera2015}, video is preferred over other preparation materials (such as podcasts and books), and its usage has been studied.
Okumoto et al. proposed the ``Response Collector'' (RC) for preparation video watching~\cite{Okumoto2018}.
The RC collects learners' responses to support classroom activities.
It allows learners to mark content as ``Interesting'', ``Important'', ``Difficult'', or add a ``Question'' on a lecture video.
The ``Question'' option provides especially important information that helps instructors to decide in-class topics because it shows in detail what learners have difficulty understanding.

Flipped learning systems are dependent on learners preparing independently before class.
Okumoto et al. suggest that some learners are not inclined to watch the video; there is a need to support these learners in flipped learning practices.
Herreid and Schiller~\cite{Herreid2012} said that learners may resist flipped learning because learners are required to work before class, framing this as a problem with these systems.
In flipped learning, if learners do not prepare before class, they will not be able to perform well in the classroom.
Clearly, additional functionality is required to ensure that video material is effective.

To make learners prepare before class, it is important to encourage proactive learning~\cite{SamsAaronJonathan2012}.
A popular method is assigning homework~\cite{DeLozier2017}; however, we believe that this may give learners a sense of compulsion and deprive them of self-motivation, thus hindering learning.

We propose a way to improve learners' motivation for preparation learning by incorporating annotations from instructors into video material.
Preparation videos are often accessed on the Internet or using DVDs~\cite{Abeysekera2015}.
However, using common methods, it is difficult for learners to understand the overall structure and contents of the learning material at a glance;
hence, they watch the videos without understanding the overall structure.
We assume that this is one of the factors that decreases their motivation to learn in preparation for classes - Yilmaz~\cite{Yilmaz2017} has shown that learners' interest in engaging with the learning material corresponds to their motivation and study satisfaction.

This paper proposes and evaluates additional mechanisms that let the instructor add marks to the flipped learning video material.
Our mechanism is realized on top of the RC, and these marks can be regarded as the first response to the video from the instructor.
These marks from the instructor show learners the overall structure and focus of the video material, and our experiments show that they improve learners’ impressions of the videos.
Therefore, the proposed mechanism will encourage more learners to watch the videos.
Our contributions are as follows.
\begin{enumerate}
  \item We argue the problem of missing learner-instructor interaction during preparation learning in flipped learning systems (in Section~\ref{sec:approach}).
  \item We propose a simple mechanism for inserting instructor marks on videos (in Section~\ref{sec:proposed_method}).
  \item We discussed that this may form a kind of interaction between instructors and learners during preparation learning (in Section~\ref{sec:approach}).
  \item We show, through experiments, that the mechanism improves impressions of the videos (in Section~\ref{sec:user_study}). 
\end{enumerate}
\section{Approach}
\label{sec:approach}
\begin{figure}[tbp]
  \centering
  \subfigure[RC without Steering Mark]{\includegraphics[width=\linewidth]{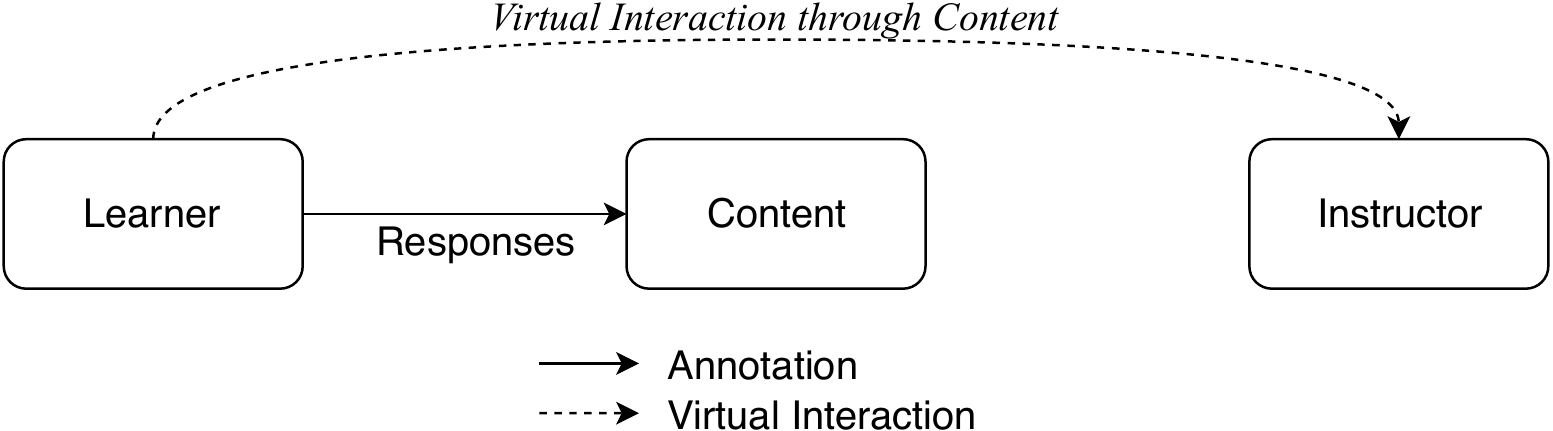}\label{fig:interactions_before}}
  \subfigure[RC with Steering Mark]{\includegraphics[width=\linewidth]{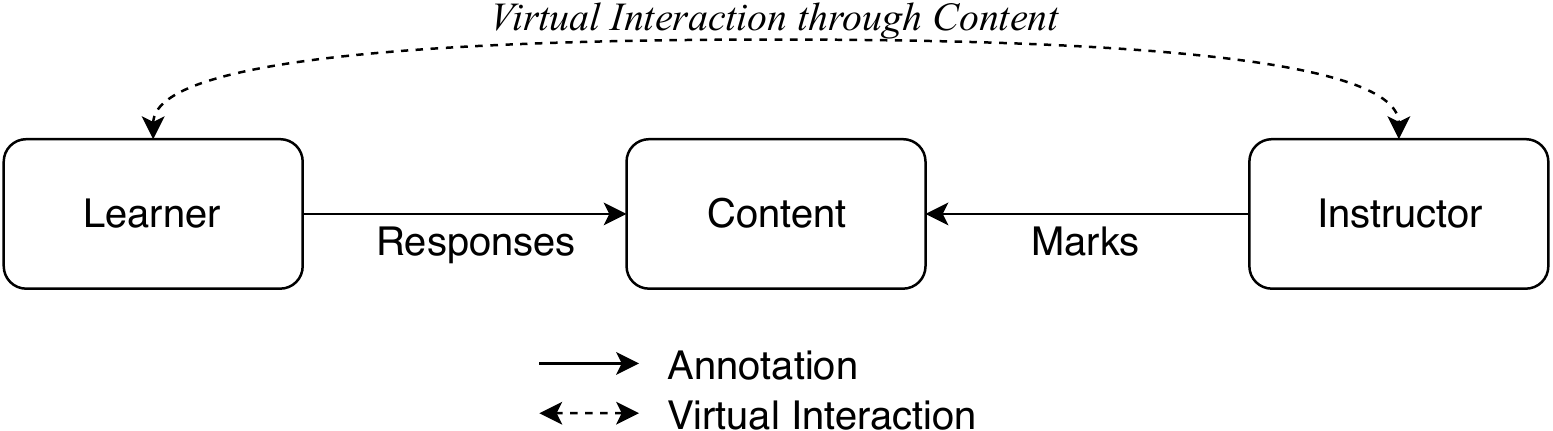}\label{fig:interactions_after}}
  \caption{By adding annotations to content, virtual interaction is formed. Introducing Steering Marks to RC, we can achieve interactions in both directions.}
  \label{fig:interactions}
  \vspace{-1em}
\end{figure}
The first step in applying flipped learning involves using a system that collects learners' responses to apply them in the classroom.
We noticed that the current RC lacks the idea of interaction between learners and instructors~\cite{Anderson2003}.
We can think preparation learning as a kind of distance education.
We should let students and instructors interact more on preparation learning.
Moreover, Kuo et al.~\cite{Kuo2014} reported that learner-content and learner-instructor interactions affect learner satisfaction.
Therefore, we expect that rather than enforcement methods such as homework assignments, improving the quality of content for learners and increasing interactions will enhance preparation learning satisfaction.
This in turn would enhance learners’ motivation to watch videos again and continue to prepare before classes.
Now, the problem lies in realizing interactions during preparation learning. 

As in Fig.~\ref{fig:interactions_before}, RC without Steering Mark collects responses from learners and provides virtual interaction between learners and instructors. We call these responses interactions because they are synchronized with the video. We have chosen the same mechanism but in reverse, as shown in Fig.~\ref{fig:interactions_after}.
First, the instructor watches the video material, considering the learners’ perspective, and they add marks and messages to the video. These will show the focus of the video as well as its overall structure.
This will enable virtual communication from instructors to learners, thus achieving interactions in both directions, as is shown in Fig.~\ref{fig:interactions_after}.

\section{Proposed Method}
\label{sec:proposed_method}
\begin{figure*}[tp]
  \vspace{-2em}
  \centering
  \subfigure[Overview. Steering Mark is indicated by a bicycle icon at the bottom.]{\includegraphics[width=0.6\linewidth]{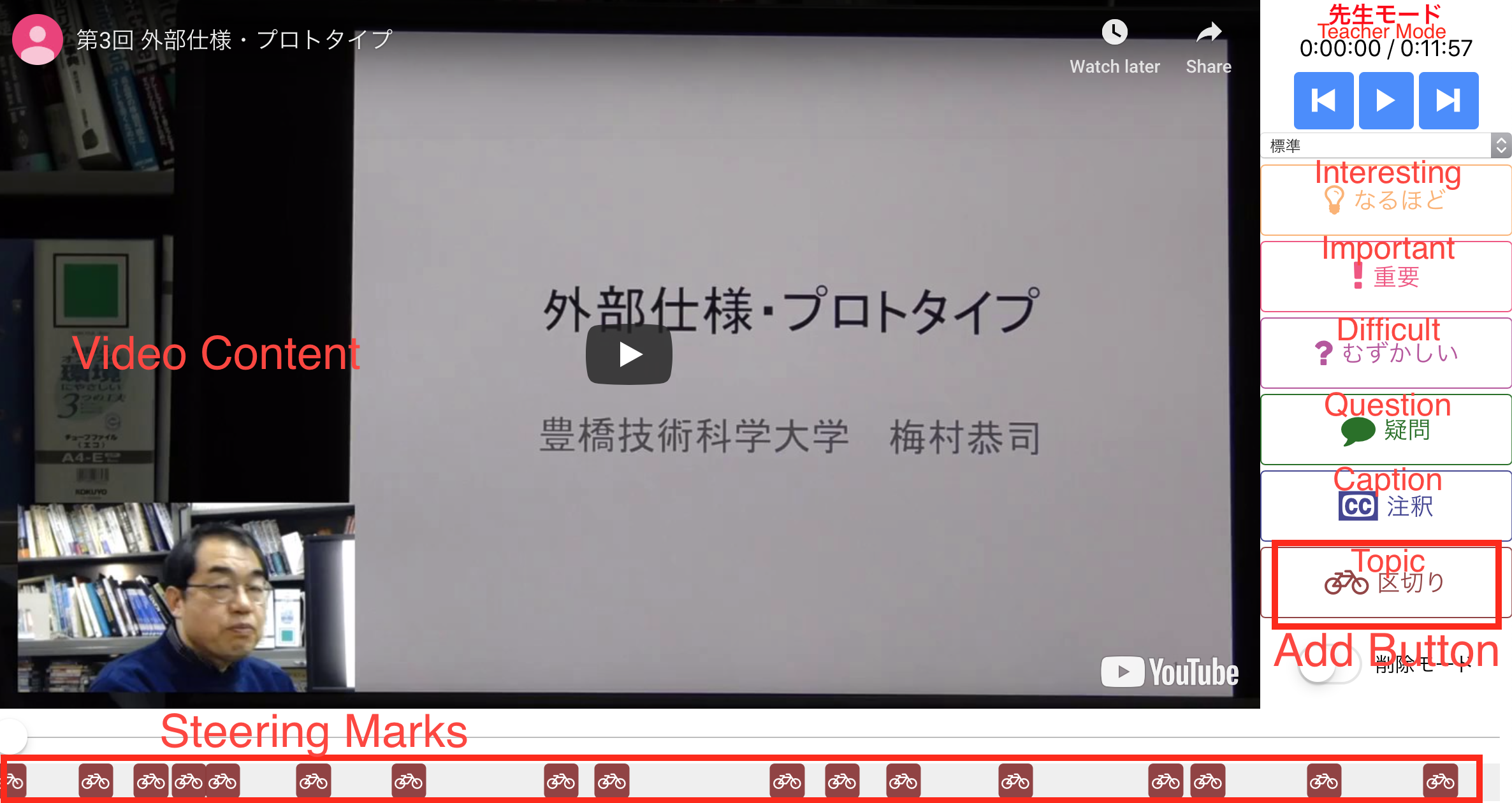}\label{fig:system_overview}}
  \subfigure[On mouseover, the title will be shown.]{\includegraphics[width=0.35\linewidth]{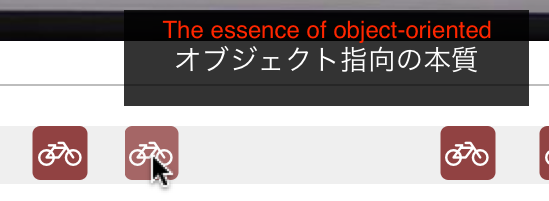}\label{fig:system_mouseover}}
  \caption{\textbf{RC with Steering Mark:} Our proposed method ``Steering Mark'' on RC. We expect that the mark lets learners grasp the overall structure of the video at a glance.}
  \label{fig:system}
  \vspace{-1em}
\end{figure*}

We incorporated ``Steering Mark'' into RC to motivate learners to prepare for classes by using a two-way interaction system.
As shown in the bottom of Fig.~\ref{fig:system_overview}, Steering Mark on RC provides learners with timestamps in the video to show topic changes.
Additionally, if a title is set to Steering Mark, it will be shown on mouseover as Fig.~\ref{fig:system_mouseover} shows.
We expected the primary advantage of Steering Mark to be an improvement in impressions of the video content because learners could grasp the structure from the start, and they could use Steering Mark to skip already known topics.

Video annotation is a well-known method for educational technology.
However, we believe that Steering Mark has difference based on flipped learning structure.
Preparation videos in flipped learning are just focused on teaching basic knowledge of lecture, unlike distance education.
Therefore, in other words, we must think the use of the knowledge in in-class activity, not just let learners watch the videos.
Thus, our methods' difference is that an instructor adds Steering Marks into the video while thinking about that learners' use of knowledge in the video.

We developed Steering Mark to work similarly to when learners add their own marks on the related system~\cite{Okumoto2018}.
An instructor can add a Steering Mark into a video as they like; click on the button to the right of Fig.~\ref{fig:system_overview} while playing the video, and a dialog that sets the title to the Steering Mark appears and the video pauses.
We assumed that this procedure would not increase instructors' labor time because they could check the video and add Steering Marks simultaneously.
\section{User Study}
\label{sec:user_study}
We set research questions to evaluate whether Steering Mark on RC works effectively to fulfill its purpose.
\begin{LaTeXdescription}
  \item[RQ(A)] Is the Steering Mark useful as a function that adds value to watching the video on the system?
  \item[RQ(B)] What kind of influence does the Steering Mark have on the collection of responses from learners?
\end{LaTeXdescription}
In order to verify these research questions, we performed experiments using a system incorporated with the Steering Mark.
\begin{table*}[tp]
  \vspace{-2em}
  \caption{Questionnaire of Impression Evaluation} \label{tbl:questions}
  \begin{center}
  \begin{tabular}{ r | p{44em} }
    No. & Question\\
     & Answer form \\
    \hline \hline
    1 & Did you watch the entire n-th video contents provided by the Response Collector?\\
      & Binary (watched the whole， watched only a part or not watched)\\
    \hline
    1-1 & Did you check the contents of the whole n-th video provided by the Response Collector while performing this exercise?\\
     & Binary (yes, no)\\
    \hline
    1-2 & Please choose the one closest to the reason you did not watch the whole video\\
     & Four alternatives (Already known/It was difficult to watch/System problem/Other)\\
    \hline
    2 & Was it difficult to watch the video provided by the Response Collector when you first watched it?\\
     & Four-step evaluation\\
    \hline
    3 & Was it easy to grasp all the content?\\
     & Four-step evaluation\\
    \hline
    4 & Did you add marks that were available?\\
     & Four-step evaluation\\
    \hline
    5 & Did the use of Steering Mark (with slide title) change?\\
     & Four-step evaluation\\
    \hline
    6 & Did the Steering Mark (with slide title) change your impression of the video content?\\
     & Four-step evaluation\\
    \hline
    7 & Was it troublesome to watch the video provided by the Response Collector when you watched it during the exercises?\\
     & Four-step evaluation\\
    \hline
    8 & How much did you refer to the content of the video provided by the Response Collector in performing the exercise?\\
     & Four-step evaluation\\
    \hline
    9 & Were you able to quickly find the topic you want to see in the video provided by the Response Collector?\\
     & Four-step evaluation\\
    \hline
    10 & Did you use the marks that you inserted?\\
     & Four-step evaluation\\
    \hline
    11 & Were the inserted marks useful?\\
     & Four-step evaluation\\
    \hline
    12 & Did you use Steering Mark (with slide title)?\\
     & Four-step evaluation\\
    \hline
    13 & Was Steering Mark (with slide title) useful?\\
     & Four-step evaluation\\
    \hline
    14 & Which did you use more - marks which you added or Steering Mark (with slide title)?\\
     & Four-step evaluation\\
    \hline
    15 & This video is distributed using YouTube. Did you feel that you wanted to watch in your familiar environment instead of the Response Collector? \\
     & Binary (yes, no)\\
    \hline
    15-1 & Please enter the reason that you prefer to watch in your familiar environment.\\
     & Free text\\
    \hline
    16 & Additional Comments\\
     & Free text
  \end{tabular}  
\end{center}
\end{table*} 
\begin{table*}[tbp]
  \vspace{-2em}
  \caption{Content and system setting in each class time} \label{tbl:settings}
  \vspace{-1em}
  \begin{center}
  \begin{tabular}{ r | c | c | c }
    Time & Content & Video length & Settings \\
    \hline \hline
    First class & Introduction of Git & 22m43s & None\\
    Second class & Git with GitHub & 15m27s & Steering Mark was added to the slide break\\
    Third class & Prototyping and external design & 11m57s & Steering Mark with slide title was added to the slide break\\
    Fourth class & Testing & 12m32s & None
  \end{tabular}
\end{center}
\end{table*}
\begin{table*}[tbp]
  \vspace{-2em}
  \caption{Number of people who answered to the questionnaire} \label{tbl:answerCounts}
  \vspace{-1em}
  \begin{center}
    \begin{tabular}{r|c|c|c|c}
      Class time & 1 & 2 & 3 & 4 \\
      \hline \hline
      Number of people & 51 & 50 & 51 & 35
    \end{tabular}
  \end{center}
  \vspace{-2em}
\end{table*}
\subsection{Experiment Procedure}
We conducted an experiment with 51 learners who took one of the classes for ``Software Exercise'' at Toyohashi University of Technology, Japan.
We encouraged learners to use the system as a reference in performing exercises after preparation learning. Table~\ref{tbl:settings} shows content and system settings in each class timeslot.
These classes are all in Japanese.

In this experiment, we set up preparation learning time using our system in the class so as to evaluate the change in impression caused by Steering Mark.
At the end of each class, we gave learners an impression evaluation questionnaire as shown in Table~\ref{tbl:questions}.
For statistical analysis, we conducted tests using a previous class as a control group.
We also used the user behavior log of the mark addition history and the playing history from the system.

In the results, we used Wilcoxon's signed rank test for the four-step evaluated questions and McNemar's test for the binary questions.
We used the paired-t test for log analyzing.
But we normalized the logs because each class used different video lengths, as shown in Table~\ref{tbl:settings}.

\subsection{Result}
Table~\ref{tbl:answerCounts} shows the number of people who answered the questionnaire.
As Table~\ref{tbl:answerCounts} shows, some learners did not answer the questionnaire, therefore, in the subsequent statistical tests we used 34 subjects who answered all the questions.

Two questions, Q6 (Fig.~\ref{fig:Q6}) and Q9 (Fig.~\ref{fig:Q9}), that ask about the relevance of Steering Mark to the video content and two questions that ask about the usage of Steering Mark versus a self-mark have significant differences in their responses.
We tested logs that counted the number of times a self-mark was added, self-mark was clicked and Steering Mark was clicked, and then we obtained the significant differences between the self-mark click log (Fig.~\ref{fig:selfMark_used}) and Steering Mark click (Fig.~\ref{fig:topicMark_used}) log.
\subsubsection{Influence of Steering Mark to the video content}
\label{sec:result1}
\begin{figure}[t]
   \begin{center}
    \includegraphics[width=\linewidth]{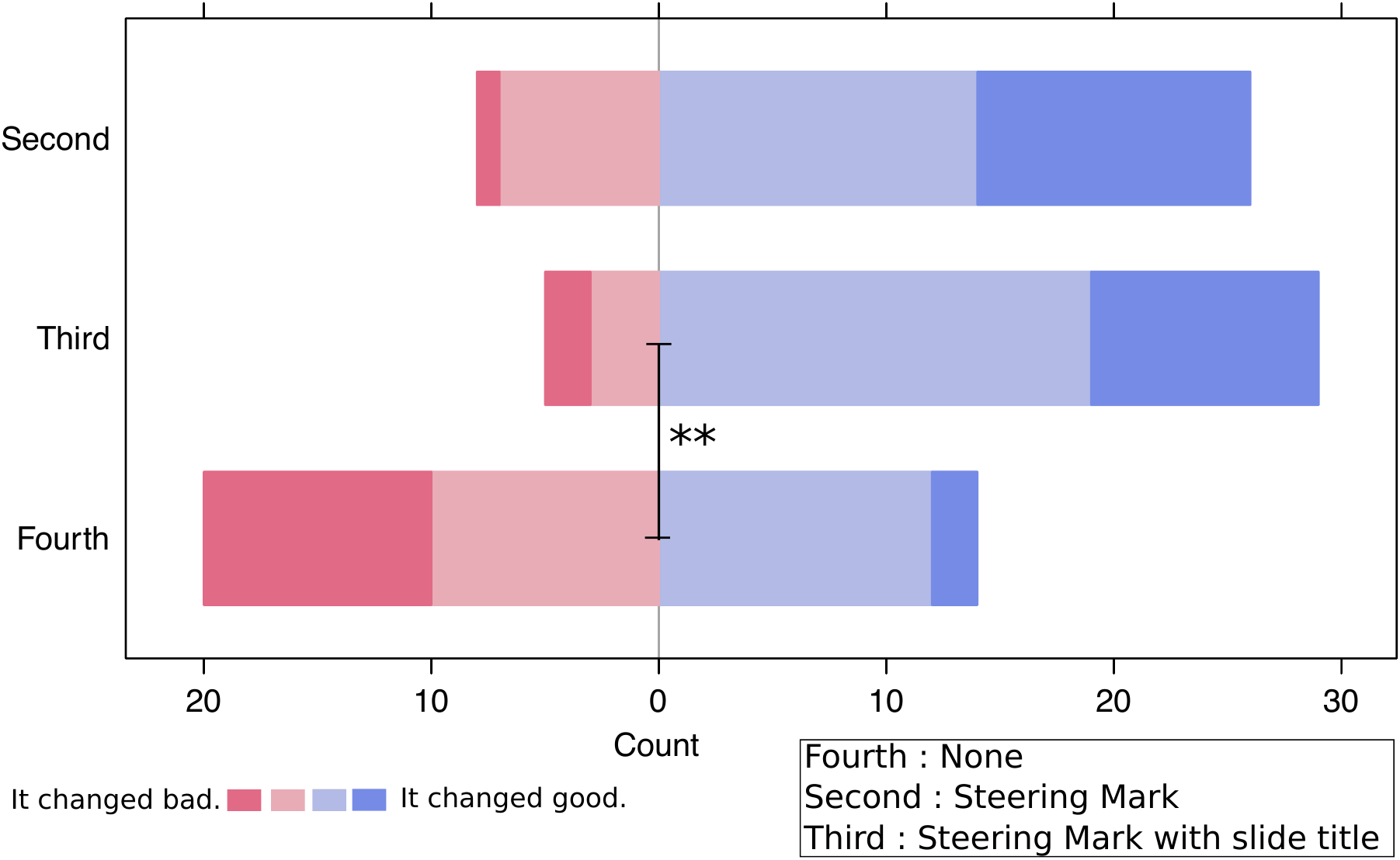}
   \end{center}
   \vspace{-1em}
   \caption{\textbf{Aggregation of Q6:} ``Did the Steering Mark (with slide title) change your impression of the video content?'' In this result, there is a significant difference in the third vs. fourth. This shows that positive impressions of the video content decreased in the fourth because Steering Mark was not active. \protect \linebreak \textit{Note. $ ^{**}p < 0.01 $}}
   \label{fig:Q6}
   \vspace{-1em}
\end{figure}
\begin{figure}[t]
   \begin{center}
    \includegraphics[width=\linewidth]{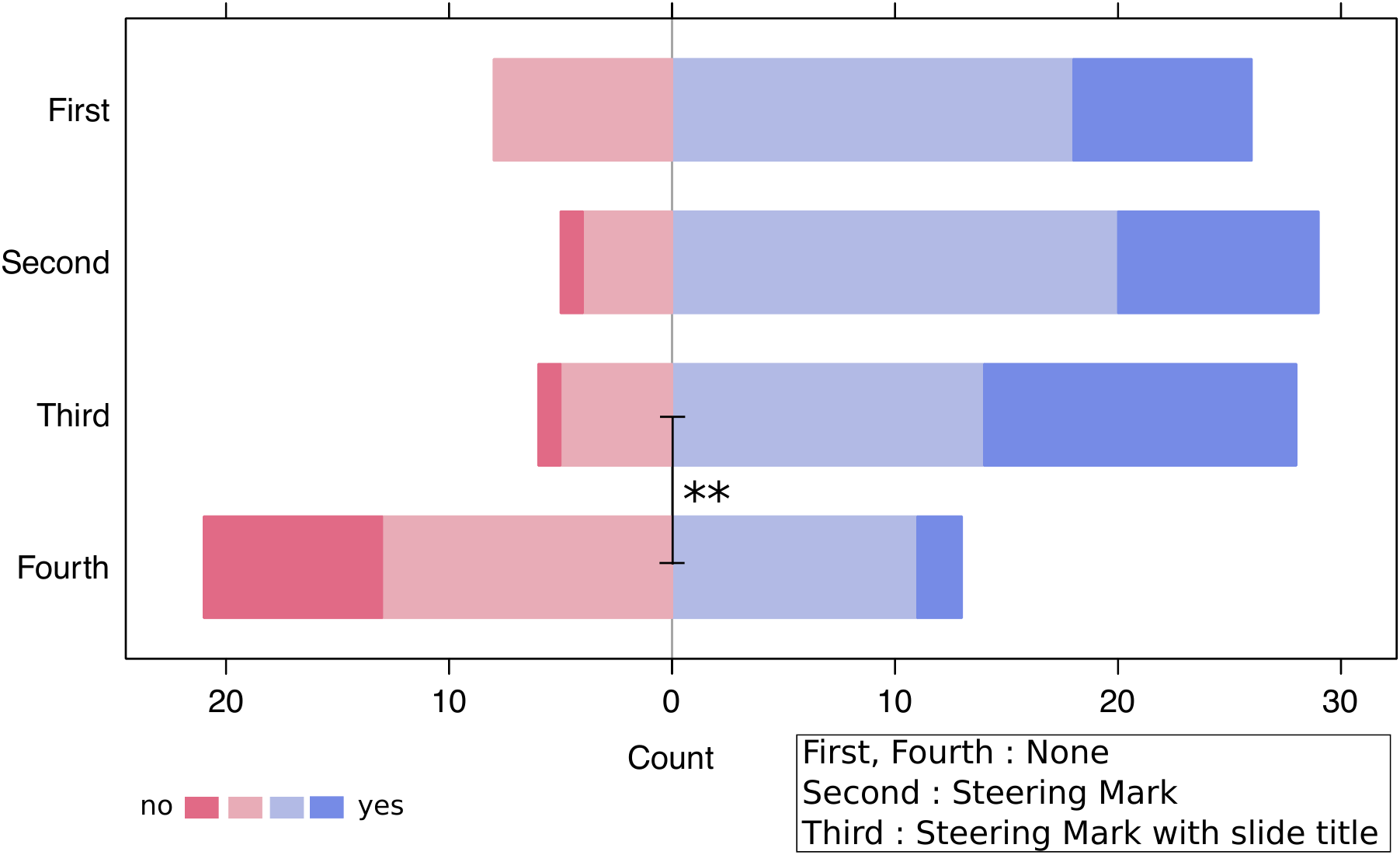}
   \end{center}
   \vspace{-1em}
   \caption{\textbf{Aggregation of Q9:} ``Were you able to quickly find the topic you want to see in the video provided by the Response Collector?'' In this result, there is a significant difference in the third vs. fourth. This shows that subjects were not able to find the topic that they wanted to watch. Additionally, 75\% or more subjects responded with ``good'' in both the second and third. \protect \linebreak \textit{Note. $ ^{**}p < 0.01 $}}
   \label{fig:Q9}
   \vspace{-1em}
\end{figure}
\begin{figure}[t]
    \begin{center}
     \includegraphics[width=\linewidth]{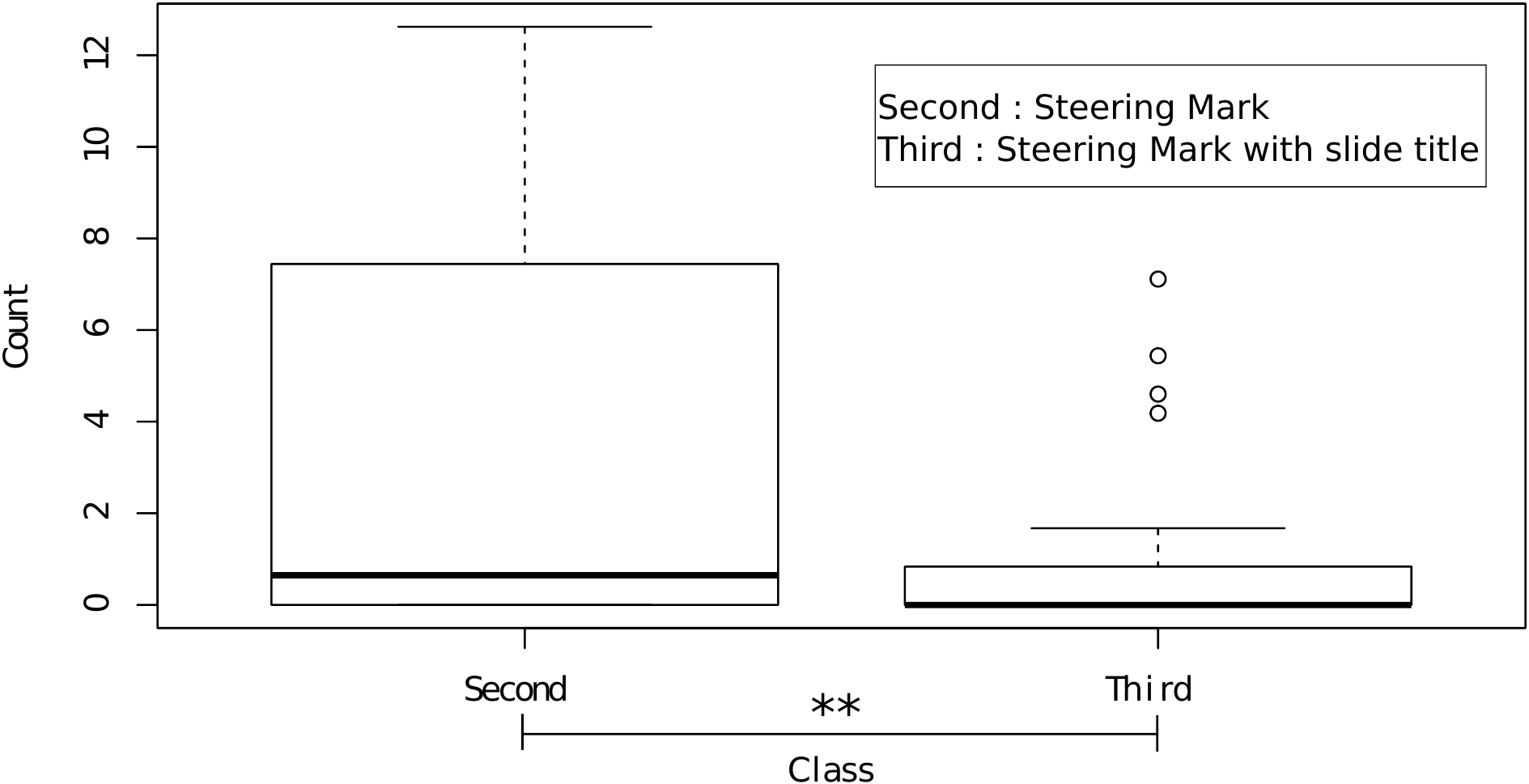}
    \end{center}
    \vspace{-1em}
    \caption{\textbf{Aggregation of Steering Mark click log per 5 min:} In this result, there is a significant difference in the second vs. third. This shows that subjects clicked Steering Mark in the second more than third. \protect \linebreak \textit{Note. $ ^{**}p < 0.01 $}}
    \label{fig:topicMark_used}
    \vspace{-1em}
\end{figure}
\begin{figure}[t]
   \begin{center}
    \includegraphics[width=\linewidth]{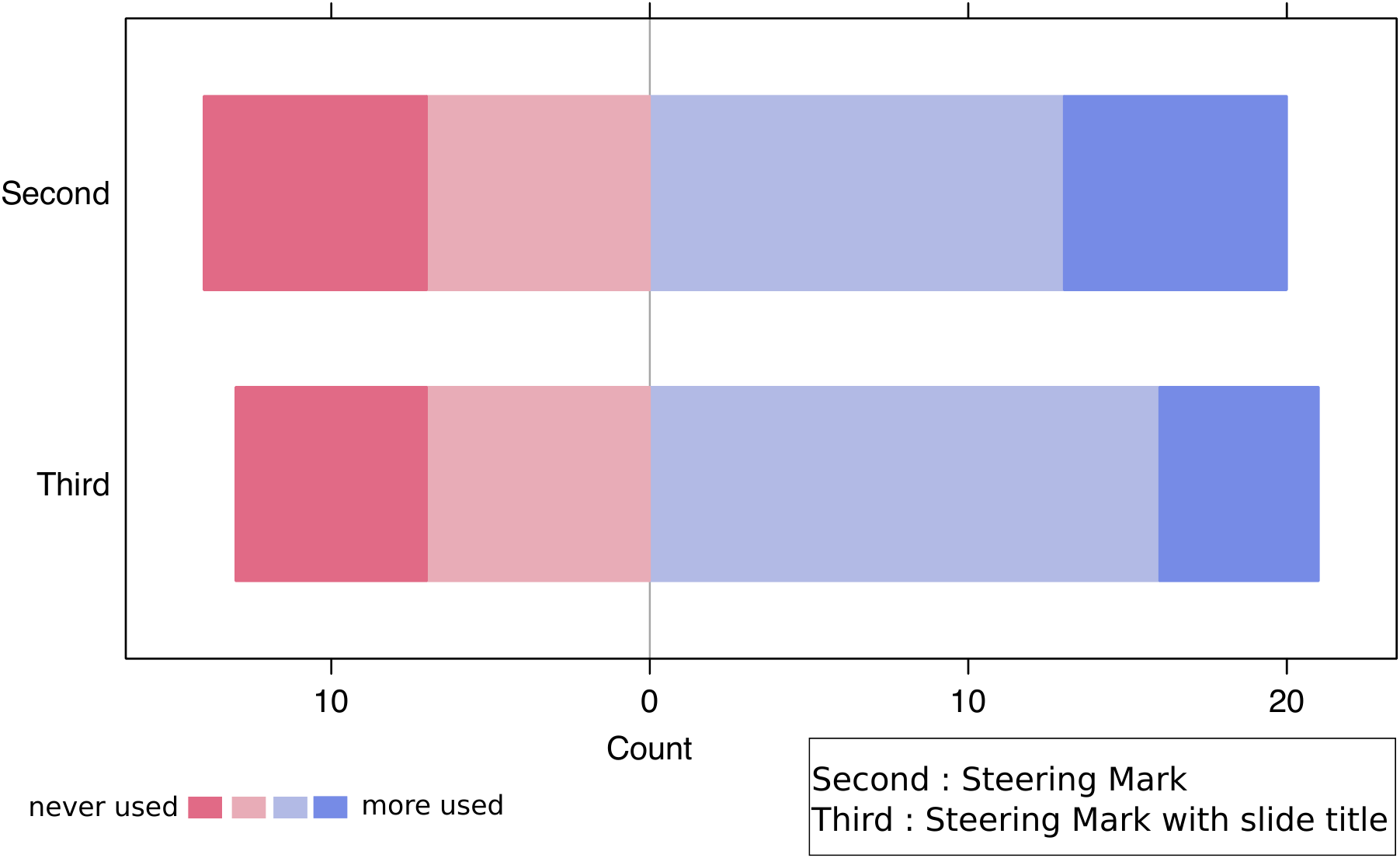}
   \end{center}
   \vspace{-1em}
   \caption{\textbf{Aggregation of Q5:} ``Did the use of Steering Mark (with slide title) change?'' In this result, there is no significant difference.}
   \label{fig:Q5}
   \vspace{-1em}
\end{figure}
We used questionnaire aggregation and log analysis to find out whether Steering Mark is useful as a function that adds value to videos to encourage watching on the system.

The questionnaire results of both Q6 - ``Did the Steering Mark (with slide title) change the impression of the video content?'' (Fig.~\ref{fig:Q6}) - and Q9 - ``Were you able to quickly find the topic you wanted to see in the video provided by the Response Collector?'' (Fig.~\ref{fig:Q9}) - show significant differences ($ p < 0.01 $) in the third class vs. the fourth class.
It was shown that the subjects' impression of the video content was negatively impacted when moving from an environment with Steering Mark to an environment without it.
Moreover, 75\% or more subjects responded with ``good'' in both the second and third classes answers to Q6 and Q9.
Additionally, the Steering Mark's click log (Fig.~\ref{fig:topicMark_used}) shows a significant difference ($ p < 0.01 $).
In this result, the mean difference is -2.86.

It was shown that the impression of the video content was improved using Steering Mark.
On the other hand, it was not possible to verify whether it had any influence on the content when slide titles were added to the Steering Mark.
\subsubsection{Influence of Steering Mark on response collection from learners}
\label{sec:result2}
We used questionnaire aggregation and log analysis to find out whether Steering Mark influenced response collection from learners.

\begin{figure}[t]
  \begin{center}
   \includegraphics[width=\linewidth]{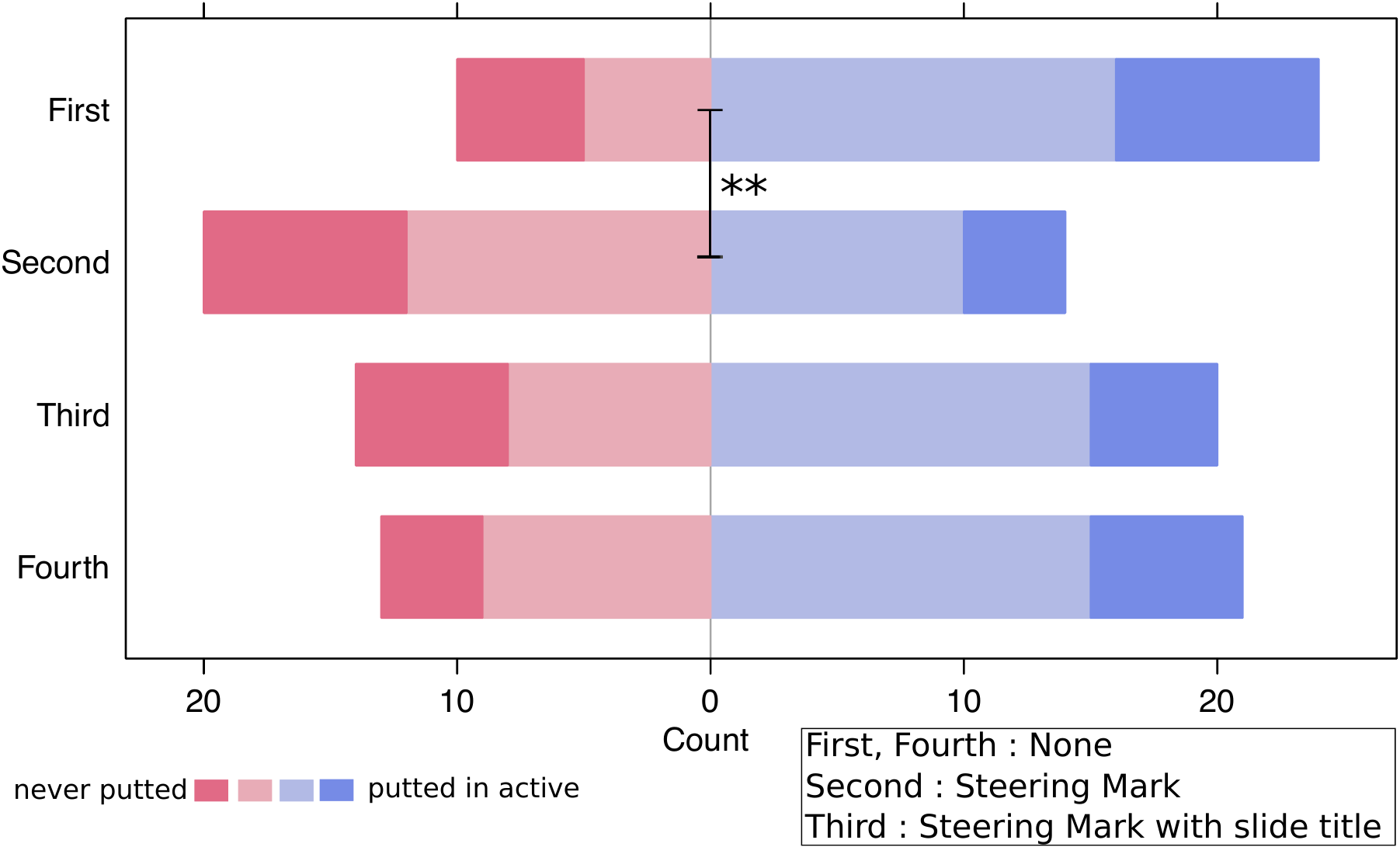}
  \end{center}
  \vspace{-1em}
  \caption{\textbf{Aggregation of Q4:} ``Did you add marks that were available?'' In this result, there is a significant difference in the first vs. second. This shows that subjects no longer put self-marks in the second. \protect \linebreak \textit{Note. $ ^{**}p < 0.01 $}}
  \label{fig:Q4}
  \vspace{-1em}
\end{figure}

\begin{figure}[t]
  \begin{center}
   \includegraphics[width=\linewidth]{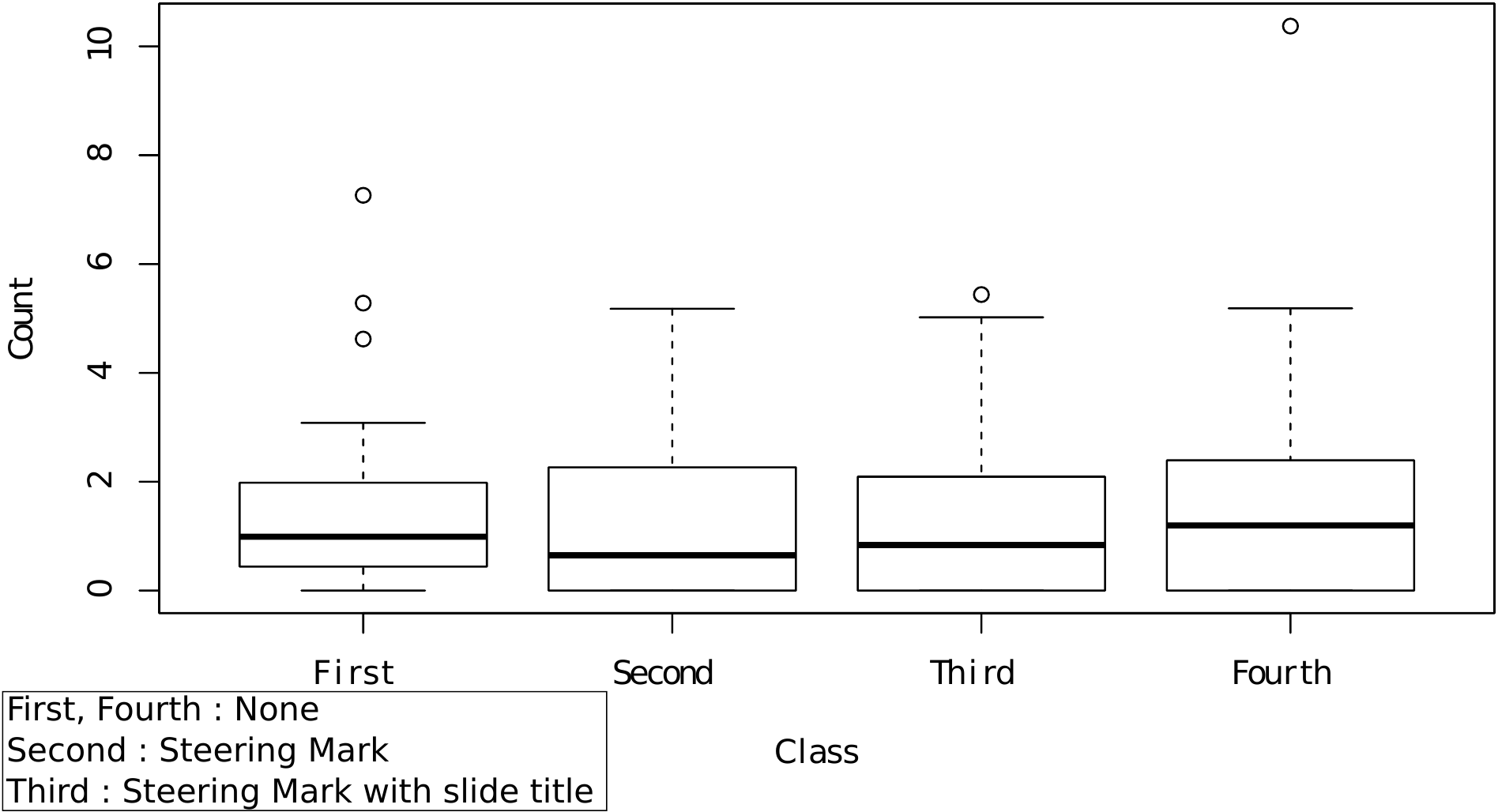}
  \end{center}
  \vspace{-1em}
  \caption{\textbf{Aggregation of self-mark added log per 5 min:} In this result, there is no significant difference.}
  \label{fig:selfMark_count}
  \vspace{-1em}
\end{figure}

\begin{figure}[t]
  \begin{center}
    \includegraphics[width=\linewidth]{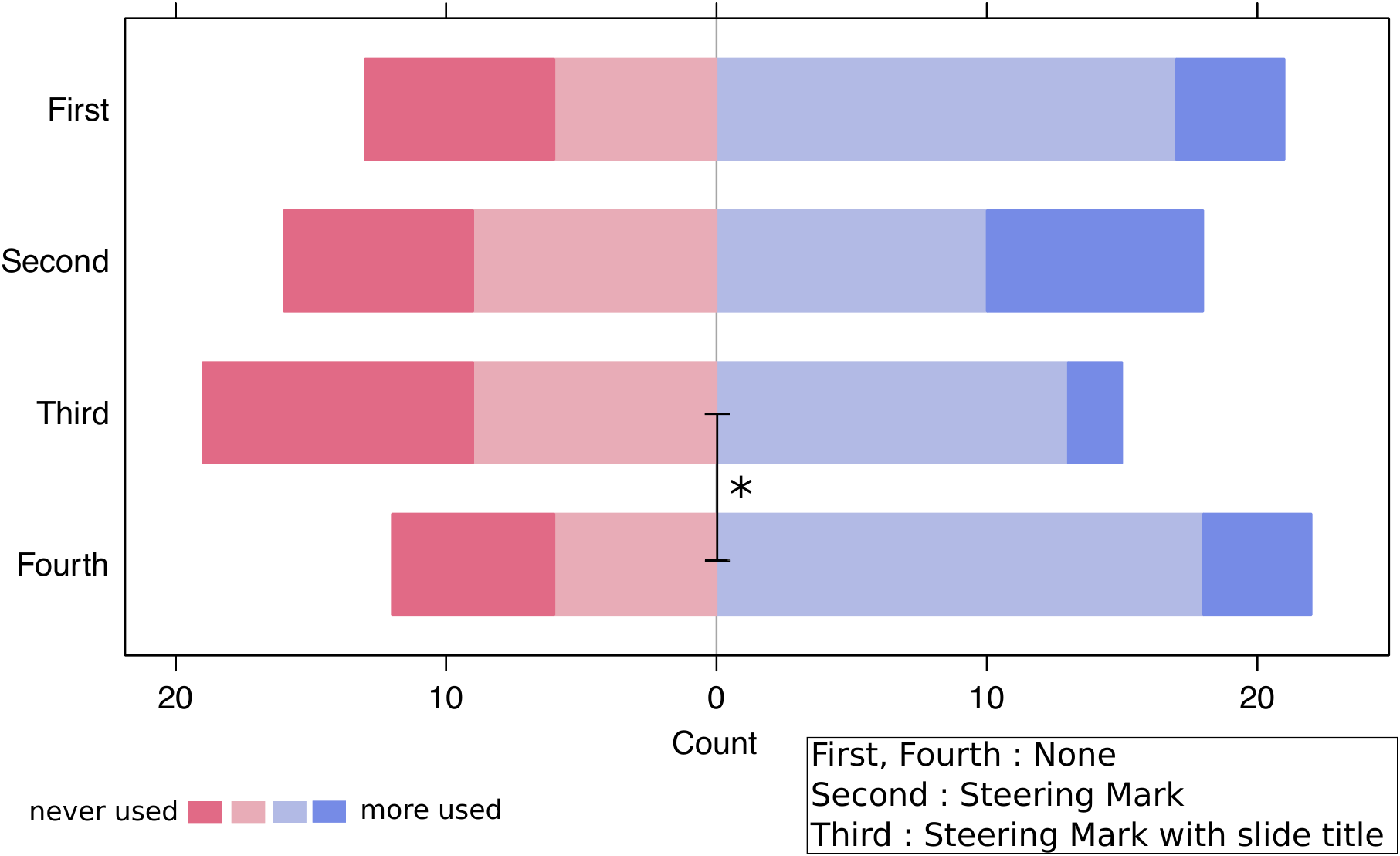}
   \end{center}
   \vspace{-1em}
   \caption{\textbf{Aggregation of Q10:} ``Did you use the marks which you added?'' In this result, there is a marginal difference in the third vs. fourth. This suggests that subjects use self-marks as alternative to Steering Mark in the fourth. \protect \linebreak \textit{Note. $ ^{*}p < 0.05 $}}
   \label{fig:Q10}
   \vspace{-1em}
\end{figure}

\begin{figure}[t]
   \begin{center}
    \includegraphics[width=\linewidth]{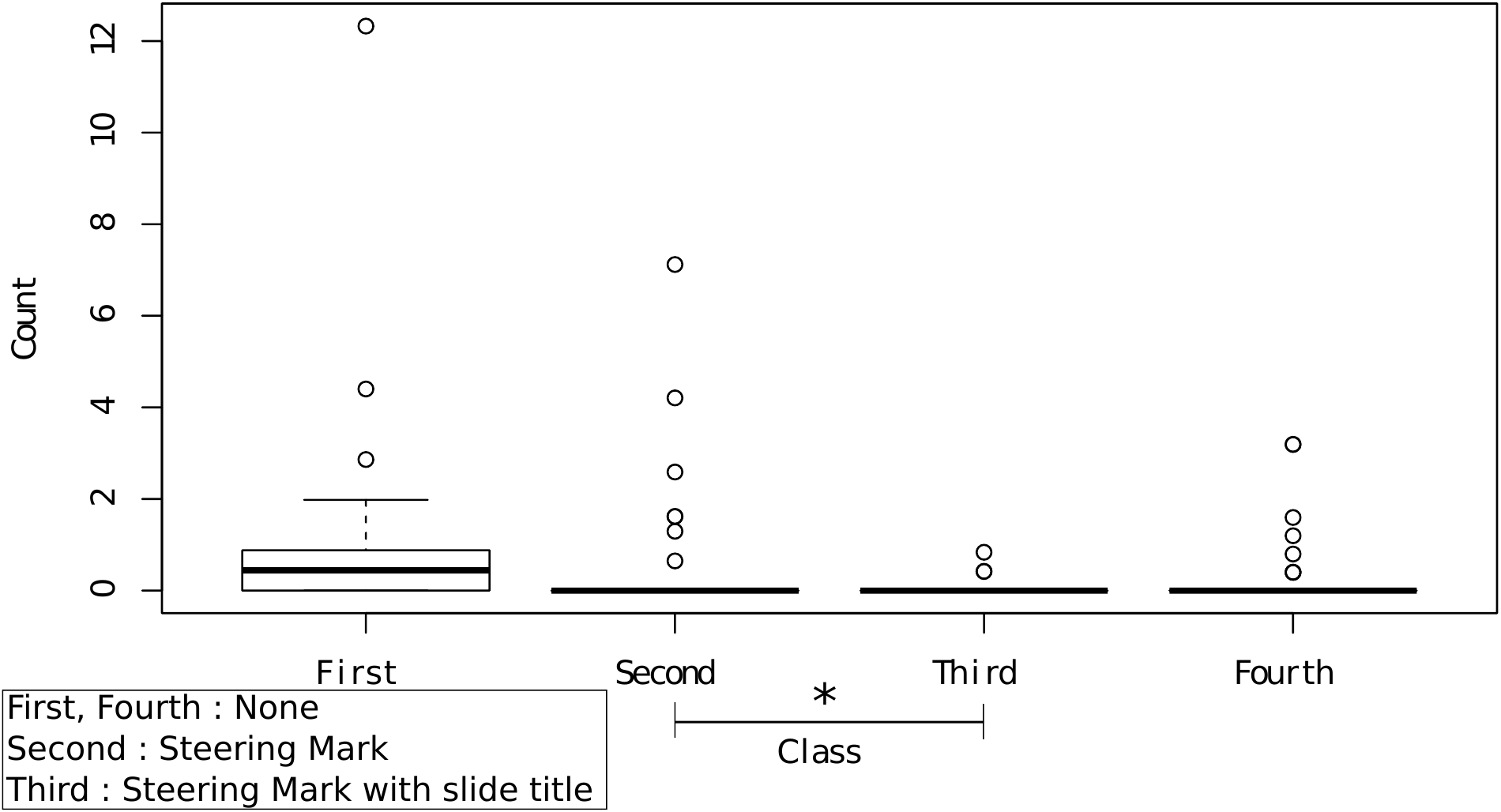}
   \end{center}
   \vspace{-1em}
   \caption{\textbf{Aggregation of self-mark click log per 5 min:} In this result, there is a marginal difference in the second vs. third. This suggests that subjects no longer click on self-marks in the third. \protect \linebreak \textit{Note. $ ^{*}p < 0.05 $}}
   \label{fig:selfMark_used}
   \vspace{-1em}
\end{figure}

A questionnaire result of Q4; ``Did you add marks that were available?'' (Fig.~\ref{fig:Q4}) shows a significant difference ($ p < 0.01 $) in the first class vs. the second class.
However, the log of self-marks added (Fig.~\ref{fig:selfMark_count}) does not show a significant difference in any set.
Alternatively, responses to Q10, ``Did you use the marks which you added?'', (Fig.~\ref{fig:Q10}) suggests ($ p < 0.05 $) that subjects clicked on self-marks more in the fourth class than the third class.
The self-mark click log (Fig.~\ref{fig:selfMark_used}) suggests ($ p < 0.05 $) that subjects click on self-marks less in the third class than the second class.
On the other hand, the log does not show or suggest significant differences in the third class vs. the fourth class, but it tends to increase ($ p < 0.1 $).

It was suggested that with the influence of Steering Mark, the use of self-mark tends to decrease for learners' response collection.
\section{Discussion}
\subsection{RQ(A):Usefulness of Steering Mark}
Our results show that Steering Mark is indeed useful as a function that adds value to video content to promote watching on the system, as shown in Section~\ref{sec:result1}.
Additionally,  we found that Steering Mark makes it easy for learners to access topics that they want to watch.
These results are supported by several answers such as ``It is easy to review when the Steering Mark is available'' and ``It is easy to access the topic that I wish to watch when the Steering Mark is available'' from the additional comments in the questionnaire.
It also shows that they may recognize the usefulness of Steering Mark.

It was unclear whether adding a slide title to Steering Mark influenced changes in learners’ impressions.
The number of Steering Mark clicks decreased significantly as Fig.~\ref{fig:topicMark_used} shows; however,  we can say that there was no difference in subjects' recognition of its use, as it is possible that  
they could find topics that they wanted to watch without clicking on the Steering Mark.
In addition, we received the following response: ``The slide title of the Steering Mark was very good''.
Therefore, when we add a slide title to the Steering Mark, it changes user behavior. 
\subsection{RQ(B):The influence of Steering Mark}
As a result of examining what kind of influence Steering Mark has on the collection of learner responses, which is the purpose of RC, it did not become statistically clear if Steering Mark is added to the video contents, as shown in Section~\ref{sec:result2}.

We measured a significant negative difference between the first and second classes for Q4 (Fig.~\ref{fig:Q4}).
However, we did not get a significant difference in the self-mark added analysis; instead, the count slightly decreased.
For this reason, subjects might have felt as if their self-marking behavior decreased because there is a 7-min difference in the video length between the first class and the second class.
In addition, in the click log analysis, an increase in self-mark clicks was not indicated in the third class vs. the fourth class; however an increase was observed in the third class vs. the fourth class for Q10 (Fig.~\ref{fig:Q10}).
Furthermore, considering that a significant negative difference is obtained in the third class vs. the fourth class in Q6 (Fig.~\ref{fig:Q6}), it is possible that the impression of the video contents worsened because subjects tended to use self-marking as an alternative to Steering Mark, and the RC without Steering Mark was unfriendly.
\section{Conclusion}
We introduced Steering Mark into the RC.
Steering Mark facilitates annotations from instructors, adding value to the video content with the aim of having learners prepare in advance through the flipped learning system.
We examined the effectiveness and the influence of Steering Mark, in terms of the number of responses, through a contrast experiment conducted in each class in a university.
In this experiment, we analyzed the impression evaluation questionnaire that was given to learners after the end of class and watching behavior logs to examine the effectiveness and influence composing RC without Steering Mark and RC with Steering Mark.

As a result, Steering Mark was found to be a useful function to improve video content.
This suggests that the Steering Mark will help learners grasp the overall structure of the video at a glance, without influencing critical response collection from learners.
\section*{Acknowledgment}
A part of this research was supported by JSPS KAKENHI Grant Number 17KO0484.
\bibliographystyle{IEEEtran}

\end{document}